\documentclass[aps, prl, reprint, groupedaddress,
  superscriptaddress, shortbibliography, notitlepage]{revtex4-1}

    \usepackage{graphicx} % Used to insert images
    \usepackage{adjustbox} % Used to constrain images to a maximum size 
    \usepackage{color} % Allow colors to be defined
    \usepackage{enumerate} % Needed for markdown enumerations to work
    \usepackage{geometry} % Used to adjust the document margins
    \usepackage{amsmath} % Equations
    \usepackage{amssymb} % Equations
    \usepackage{eurosym} % defines \euro
    \usepackage[mathletters]{ucs} % Extended unicode (utf-8) support
    \usepackage[utf8x]{inputenc} % Allow utf-8 characters in the tex document
    \usepackage{fancyvrb} % verbatim replacement that allows latex
    \usepackage{grffile} % extends the file name processing of package graphics 
    \usepackage{hyperref}
    \usepackage{booktabs}  % table support for pandoc > 1.12.2
    \usepackage[lofdepth,lotdepth]{subfig}

    \definecolor{darkorange}{rgb}{.71,0.21,0.01}
    \definecolor{darkgreen}{rgb}{.12,.54,.11}
    \definecolor{blue}{rgb}{0,.145,.698}

    \hypersetup{
      breaklinks=true,  % so long urls are correctly broken across lines
      colorlinks=true,
      urlcolor=blue,
      linkcolor=darkorange,
      citecolor=darkgreen,
      }
\begin{document}

\title{Hamiltonian transformation to compute Thermo-osmotic Forces}
\author{Raman Ganti}
\affiliation{Department of Chemistry, University of Cambridge, Lensfield Road, Cambridge CB2 1EW, UK}
\author{Yawei Liu}
\affiliation{Beijing University of Chemical Technology, Beijing, P. R. China}
\author{Daan Frenkel}
\thanks{Corresponding author}
\email{df246@cam.ac.uk}
\affiliation{Department of Chemistry, University of Cambridge, Lensfield Road, Cambridge CB2 1EW, UK}
	
\date{\today}

\begin{abstract}

If a thermal gradient is applied along a fluid-solid interface, the fluid experiences a  thermo-osmotic force. In steady state this force is balanced by the gradient of the shear stress. Surprisingly, there appears to be no unique microscopic expression that can be used for computing the magnitude of the thermo-osmotic force. 

Here we report how, by treating the mass $M$ of the fluid particles as a tensor in the Hamiltonian, we can eliminate the balancing shear force in a non-equilibrium simulation and therefore compute the thermo-osmotic force at simple solid-fluid interfaces.  We compare the non-equilibrium force measurement with estimates of the thermo-osmotic force based on computing gradients of the stress tensor. We find that the thermo-osmotic force  as measured in our simulations cannot be derived from the most common microscopic definitions of the stress tensor. 
\end{abstract}
 
\maketitle
Nanotechnology is not just conventional technology scaled down to the nano scale. The reason is that processes that are relatively unimportant on macroscopic scales may become dominant on the nano-scale.  Case in point are phoretic flows: the movement of fluids under the influence of gradients of thermodynamic quantities such as temperature or chemical potential.  On a macroscopic scale, the application of a pressure gradient or a body force is the most efficient way to move fluid through a tube. The resulting flux is proportional to the fourth power of the tube diameter. However, on a sub-micron scale, phoretic flows tend to become important because the resulting volumetric flow rates scale as the square of the tube diameter.  Hence, for many problems, be they technological (e.g. nano-fluidics) or natural (e.g. fluid flow through porous networks or gels), it is becoming increasingly important to be able to predict phoretic flows. 

A key feature of phoretic flows is that they are driven by forces that only act on those parts of the fluid that interact with the confining surfaces. The range of the fluid-wall interactions is typically in the nano-meter regime, except in the case of electrolytes in contact with charged surfaces, in which case the interaction layers may have  thicknesses ranging from nanometers to microns.   Here we will be considering thermo-osmotic flows in non-polar fluids near a wall. For such systems, the thermo-osmotic force driving the flow is typically confined to an interfacial layer with a thickness of a few molecular diameters. Thermo-osmotic flows have been known for well over a century~\cite{lippmann1907,aubert1912}, but the relevance of this phenomenon is increasing as more experiments probe transport on the nano-scale. Moreover, there is increasing evidence that large temperature gradients may exist inside eukaryotic cells~\cite{chretien2017mitochondria}, which is also an environment full of interfaces. 
%Mitochondria Are Physiologically Maintained At Close To 50 C Dominique Chretien, Paule Benit, Hyung-Ho Ha, Susanne Keipert, Riyad El-Khoury, Young-Tae Chang, Martin Jastroch, Howard Jacobs, Pierre Rustin, Malgorzata Rak, doi: https://doi.org/10.1101/133223

Derjaguin~\cite{derjaguin_surface} formulated a generic description of thermo-osmosis in the language of irreversible thermodynamics. As the approach by Derjaguin (and others) is phrased in the language of  macroscopic thermodynamics and continuum hydrodynamics (creeping-flow equations), it cannot be used  for a quantitative prediction of the magnitude of thermo-osmotic flows from knowledge of the intermolecular interactions. Moreover, the validity of continuum hydrodynamics is questionable  in the first few molecular layers near a wall. 

Here we use molecular simulations to predict the strength of thermo-osmotic flows.  The most straightforward approach is to carry out non-equilibrium Molecular Dynamics simulations to probe thermally induced flows. Below, we will indeed describe such simulations. However, this direct approach has practical drawbacks, as a constant temperature gradient is incompatible with the periodic boundary conditions that are commonly used in simulations to minimise finite-size effects. An alternative route is inspired by the approach of Derjaguin, who used the Onsager reciprocity relations to relate the flow due to a temperature gradient to the more easily calculated, excess heat transport due to a pressure gradient. In fact, in earlier work~\cite{ganti2017molecular} we found reasonable agreement between the Onsager approach and non-equilibrium simulations. However, neither approach allows us to compute directly the forces on a fluid due to thermal gradients parallel to a surface. 

To gain insight on the microscopic origins of thermo-osmotic flow, it is necessary to isolate the forces due to the thermal gradients from those due to gradients in the shear stress. The normal route to obtain the force $f^V_x(z)$ on a volume element in a liquid is to compute the gradient of stress acting on that element. Here $z$ denotes the distance from the interface, and $x$ the direction of the force parallel to the wall. The superscript $V$ indicates that $f^V_x(z)$ is the force per unit volume. We can convert $f^V_x(z)$ into $f^P_x(z)$, the force per particle, by using $\rho(z)f^P_x(z) = f^V_x(z)$, where $\rho(z)$ is the number density at a distance $z$ from the wall. 
 
We note that the stress $\sigma_{xx}$ depends on $x$ only through its (explicit or implicit) dependence on temperature:
\begin{equation}
\label{eq:stress_gradient}
\frac{\partial \sigma_{xx}(z)}{\partial x} = \left(\frac{\partial \sigma_{xx}(z)}{\partial T}\right)_{P_{\rm bulk}}\frac{\partial T}{\partial x} \;.
\end{equation}
The temperature derivative is computed at constant bulk pressure because thermal gradients do not cause pressure gradients in the bulk of the liquid.  Eqn.~\ref{eq:stress_gradient} provides a convenient route to compute stress gradients numerically, because the temperature dependence of the stress tensor can be computed  from equilibrium simulations at slightly different temperatures by numerical differentiation:
\begin{equation}
\label{eq:dP_dT}
\frac{\partial \sigma_{xx}(z)}{\partial T} = \frac{\sigma^{eq,T_2}_{xx}(z) - \sigma^{eq,T_1}_{xx}(z)}{T_{2} - T_{1}}.
\end{equation}
In what follows, we denote the approach based on Eqns.~\ref{eq:stress_gradient} and \ref{eq:dP_dT} as the ``stress-gradient'' route. The stress-gradient method would seem to offer a route to compute phoretic forces in thin layers from the microscopic definition of the stress tensor. However, as we show below, this approach fails. We recall that the definition of the microscopic stress tensor is not unique. This ambiguity is not a problem when computing the bulk pressure, or even the surface tension~\cite{schofield1982statistical}. However, for stress gradients parallel to a surface, different definitions of the microscopic stress do not yield the same answer. The obvious question is then: which stress tensor provides the correct description. The surprising answer that we find is ``none'' (at least not one of the usual suspects). 

As an alternative to computing the gradient of the microscopic stress tensor, we can relate the gradient of the position-dependent stress to the local value of the excess enthalpy~\cite{ganti2017molecular}:
\begin{equation}
\label{eq:gradP_enthalpy}
\frac{\partial \sigma_{xx}(z)}{\partial x} = -\left(\frac{\Delta{h(z)}}{T}\right)\frac{\partial T}{\partial x}.
\end{equation}
where $\Delta{h(z)}$ is the difference between the molar enthalpy $h(z)$ at a distance $z$ from the surface, and the bulk molar enthalpy. Note that Eqn.~\ref{eq:gradP_enthalpy} can be obtained from a single equilibrium simulation.

Once we have evaluated the gradient of the stress, the thermo-osmotic force per particle is given by
\begin{equation}
\label{eq:body_force}
f^{P}_{x}(z) =\frac{1}{\rho_{ave}(z)}\left( \frac{\partial \sigma_{xx}(z)}{\partial T} \frac{\partial T}{\partial x}\right)
\end{equation}
where $\rho_{ave}(z) = (\rho(T_{1}, P, z) + \rho(T_{2}, P, z))/2$. 

Due to the non-uniqueness of the definition of the stress tensor, different definitions may yield different stress gradients.  The most commonly used microscopic stress definitions are the virial ($V$) stress (see e.g.~\cite{hansen1990theory}) and the Irving-Kirkwood ($IK$) stress~\cite{irving1950}. Both stress definitions have identical kinetic ($K$) contributions ($\sigma^{K}_{xx}(z) = -\rho(z)k_{B}T$).  The non-uniqueness of the stress arises from different definitions of the potential ($\phi$) stress.

The $V$ stress considers the force on an atom $i$ located at position $z$ interacting via pairwise forces with neighboring atoms $j$
\begin{equation}
\label{eq:virial_pressure}
\sigma^{V,\phi}_{xx}(z) = \frac{1}{2V(z)}\sum^{N(z)}_{i} \sum_{j\ne i}\left\langle\frac{x_{ij}^{2}}{r_{ij}}\phi'(r_{ij})\right\rangle
\end{equation}
where $r_{ij}$ is the distance between atoms $i$ and $j$, $x_{ij}$ is the $x-$distance, $\phi(r_{ij})$ is the interaction potential, $V(z)$ and $N(z)$ are the bin volume and number of atoms at position $z$. 

The $IK$ stress computes the total momentum flux that crosses a fictitious surface at position $z$ in the fluid.
\begin{align}
\label{eq:irving_kirkwood}
\sigma^{IK, \phi}_{xx}(z) = \nonumber \frac{1}{2A}\sum^{N}_{j}\sum_{k\ne j} \left\langle\frac{x_{jk}^{2}}{r_{jk}}\frac{\phi'(r_{jk})}{|z_{jk}|} \right. \\ 
\times \left. \Theta\left(\frac{z - z_{j}}{z_{jk}}\right)\Theta\left(\frac{z_{k} - z}{z_{jk}}\right)\right\rangle.
\end{align}
In the $IK$ expression, the potential force that atom $j$ exerts on $k$ contributes to the stress at all points $z$ located between $z_{j}$ and $z_{k}$. 

In what follows, we assume that the structure of the confining solid does not depend on temperature.  Symmetry then implies that, on average, a flat solid wall exerts zero net transverse force on a fluid atom. However, the force derived from the gradient of the stress tensor can be non-zero, even for a flat wall. 

In addition to the stress-gradient approaches discussed above, the thermo-osmotic force can be computed using the right-hand side of Eq~\eqref{eq:gradP_enthalpy}. This `Local Thermal Equilibrium' (LTE) approach considers the following expression for the local specific enthalpy 
\begin{align}
\label{eq:local_enthalpy}
h(z) = u(z) + \frac{P_{xx}(z)}{\rho(z)} \;,
\end{align}
where $u$ is the specific internal energy. In the expression of the local enthalpy, we define the local pressure as the $xx$-component of the pressure tensor. This definition may seem arbitrary, but it follows from the Onsager relation between the excess heat flux and the thermocapillary flow: in steady state, the excess heat flux contains a term  proportional to the $xx$ component of the (virial) pressure. 
The LTE expression for the  thermo-osmotic  force per particle is given by
\begin{equation}
\label{eq:LTE_force}
f^{P}_{x}(z) =-\left( \frac{h(z) - h^{B}}{T} \right)\frac{\partial T}{\partial x}
\end{equation}
where $h^{B}$ is the bulk specific enthalpy.

\begin{figure}[tb]
\begin{center}
\includegraphics[width=1.0\columnwidth]{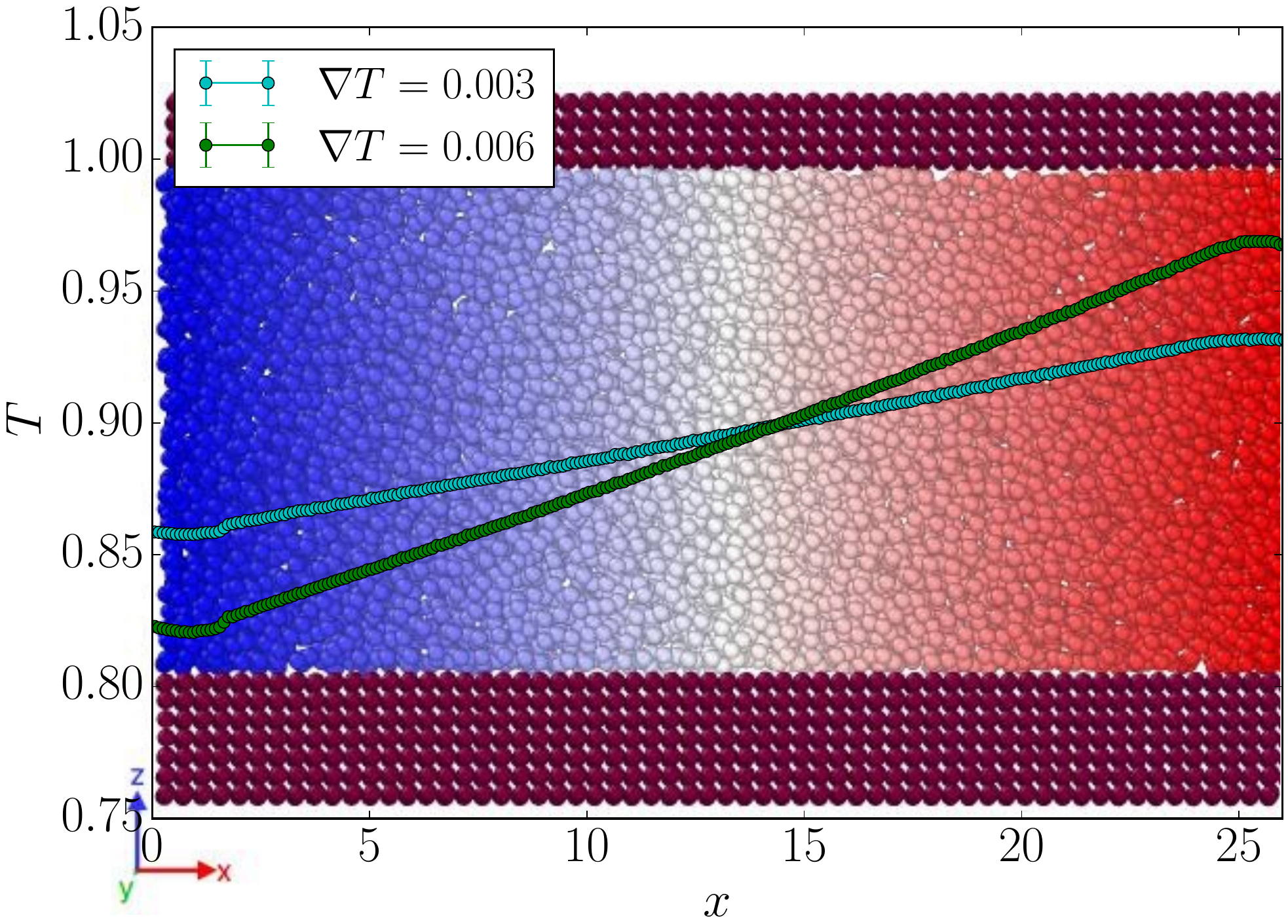}
\caption{Simulation box used for non-equilibrium force calculation where fluid near the bottom interacts with a structured wall. Temperature profiles for the simulation are plotted over the box.
%$\Delta x_{1}$ and $\Delta x_{2}$ are local equilibrium reservoirs at $(\rho_{1}, T_{1}, P)$ and $(\rho_{2}, T_{2}, P)$ respectively. 
\label{fig:simulation_box}}
\end{center}
\end{figure}

We now have at least three distinct expressions for the thermo-osmotic force and we would like to know which, if any of these, is correct. The obvious approach would be to compute the thermo-osmotic force in a steady-state, non-equilibrium simulation. However, such an approach cannot work, because in steady-state the average force on all fluid particles must necessarily vanish: the flow induced by the temperature gradient causes a gradient in shear stress that cancels the thermo-osmotic force (Eq~\eqref{eq:body_force}).

To eliminate the shear stress in a non-equilibrium simulation, such that only the thermo-osmotic force remains, we propose the following non-equilibrium simulation technique: First, we impose a periodic temperature gradient along $x$. This is done by selecting the left-most part of the simulation box  (see Fig~S2 in Supplementary Material) and thermostatting at a temperature lower than the average ($T=0.9$) while also selecting the middle of the simulation box and thermostatting at a temperature higher than the average (see Supplementary Material: Non-equilibrium Method). The resulting heat current sets up the thermal gradient. 

After the system has reached steady-state, we change the equations of motion for the fluid atoms: in  particular, we now treat the mass $M$ of the fluid particles as a tensor in the Hamiltonian, and consider the limit where $M_{yy}=M_{zz}=M$, the original mass of the particles, whilst $M_{xx}\rightarrow\infty$. Transforming the Hamiltonian in this way changes the dynamics of the system, but static properties such as inter-molecular interactions remain the same. As the temperature remains finite,  $v_{x}\rightarrow 0$ for all fluid atoms. In other words, we have switched off the shear flow, whilst maintaining the temperature gradient (see Supplementary Material: Non-equilibrium Method). 

However, fluid atoms are still diffusing in the $y$ and $z$ directions. In equilibrium, equipartition would still hold in this model system: hence, the average kinetic energy associated with motion in the $x$ direction is still $k_BT/2$. In the system with imposed thermal gradients, the magnitude of the temperature gradient is left unchanged (Fig~\ref{fig:simulation_box}). In this stationary system, the bulk serves as a reservoir of atoms so that fluid near the surface can rearrange to the local-equilibrium density profile. As the gradient in shear stress $\partial \sigma_{xz}(z)/\partial z$ now vanishes, only the thermo-osmotic force will remain. 

To compute the average thermo-osmotic force, we must average the force calculation over many different initial configurations, as every single realization will have a different density-profile in the $x$-direction frozen in.

In our direct non-equilibrium measurement, we consider a Lennard-Jones fluid consisting of $N = 7920$ atoms interacting via a truncated and shifted Lennard-Jones potential
\begin{equation}
\label{eq:lennard_jones}
V_{\mathrm{trunc}}(r)=\begin{cases}
    4\epsilon \left[\left(\frac{\sigma}{r}\right)^{12} - \left(\frac{\sigma}{r}\right)^{6}\right] - V(r_{c}) &r \le r_{c}\\
    0 &r > r_{c}.
  \end{cases}
\end{equation}
where $r_c = 4\sigma$. In what follows $\sigma$ is our unit of length and $\epsilon$ is our unit of energy:  all computed quantities are expressed in reduced units. We carried out simulations where this fluid was in contact with three different surfaces: a structured wall interacting with fluid through a less attractive Lennard-Jones potential, a structured wall interacting via a purely repulsive Weeks-Chandler-Andersen (WCA) potential ~\cite{weeks1971role}, and a reflecting wall that simply flips the corresponding velocity of fluid atoms if they attempt to cross it.
The parameters for interaction between the fluid and structured wall are: 
$ \sigma_{\text{fluid-fluid}} = \sigma_{\text{solid-fluid}} = \sigma$. The interaction strength between the fluid and structured wall is given by $\epsilon_{\text{solid-fluid}} = 0.55\epsilon$. The WCA interaction between the fluid and repulsive wall atoms was obtained by truncating and shifting the fluid-fluid interaction  at  $r_c = 2^{1/6}\sigma$.

All Molecular Dynamics simulations were carried out using the LAMMPS package~\cite{plimpton1995fast}. Fig~\ref{fig:simulation_box} shows a simulation cell of length $\langle L_{x} \rangle = 49.32\sigma$ and $\langle L_{y} \rangle = 9.86\sigma$ containing fluid that interacts with a structured wall. Fig~S1(b) shows the simulation cell for fluid interacting with a reflecting wall. To ensure that $P = 0.122$ in the bulk, the top wall acts as a piston that is free to move in the $x$ and $z$-directions.

The solid atoms in the structured walls were arranged in an fcc lattice ($\rho = 0.9 \sigma^{-3}$) bonded via harmonic springs to their nearest neighbors, where the spring stiffness $k_{bond} = 5000 \epsilon/\sigma^{2}$ and equilibrium rest length is $1.1626 \sigma$. The fluid was in contact with the $\{001\}$ face of the crystal lattice.

Using a smaller simulation box ($\langle L_{x} \rangle/3 = 16.44\sigma$, $N = 2640$ fluid atoms), the V (Eq~\eqref{eq:virial_pressure}) and IK (Eq~\eqref{eq:irving_kirkwood}) stress profiles were computed for systems at $(T = 0.85, P = 0.122)$ and $(T = 0.95, P = 0.122)$ (see Supplementary Information: Stress-Gradient Method). With $\rho(z) = (\rho(T = 0.85, P = 0.122, z) + \rho(T = 0.95, P = 0.122, z))/2$ and Eq~\eqref{eq:dP_dT}, the $V$ and $IK$ force per particle (Eq~\eqref{eq:body_force}) were computed for the temperature gradients shown in Fig~\ref{fig:simulation_box}. For the LTE approach, the specific potential energy and virial stress per atom profiles were computed for a system at $(T = 0.9, P = 0.122)$. Since at constant temperature the specific kinetic energy is uniform at all points, it is straightforward to compute the local enthalpy via Eq~\eqref{eq:local_enthalpy} and therefore the body force (Eq~\eqref{eq:LTE_force}).

When comparing the directly computed thermo-osmotic force in the non-equilibrium simulation with the `stress gradient' and LTE methods, we should note that the direct calculation will only include the gradient in the potential stress
\begin{equation}
\label{eq:potential_force}
f_{x}^{P,\phi}(z) = \frac{1}{\rho(z)}\left(\frac{\partial \sigma^{\phi}_{xx}(z)}{\partial T}\frac{\partial T}{\partial x}\right)
\end{equation}
since the force computation will simply be a summation over all pairwise forces. Yet, as mentioned previously,
the non-uniqueness of the stress arises due to different definitions of the $\emph{potential}$ stress not the kinetic. Therefore, we can use our equilibrium measurements of the kinetic stress at different temperatures (see Supplementary Information: Stress-Gradient Method) to calculate the gradient of the kinetic stress. Adding the kinetic contribution to our direct calculation should give the full thermo-osmotic force.
 
\begin{figure*}[tb]
\begin{center}
\includegraphics[width=1.0\textwidth]{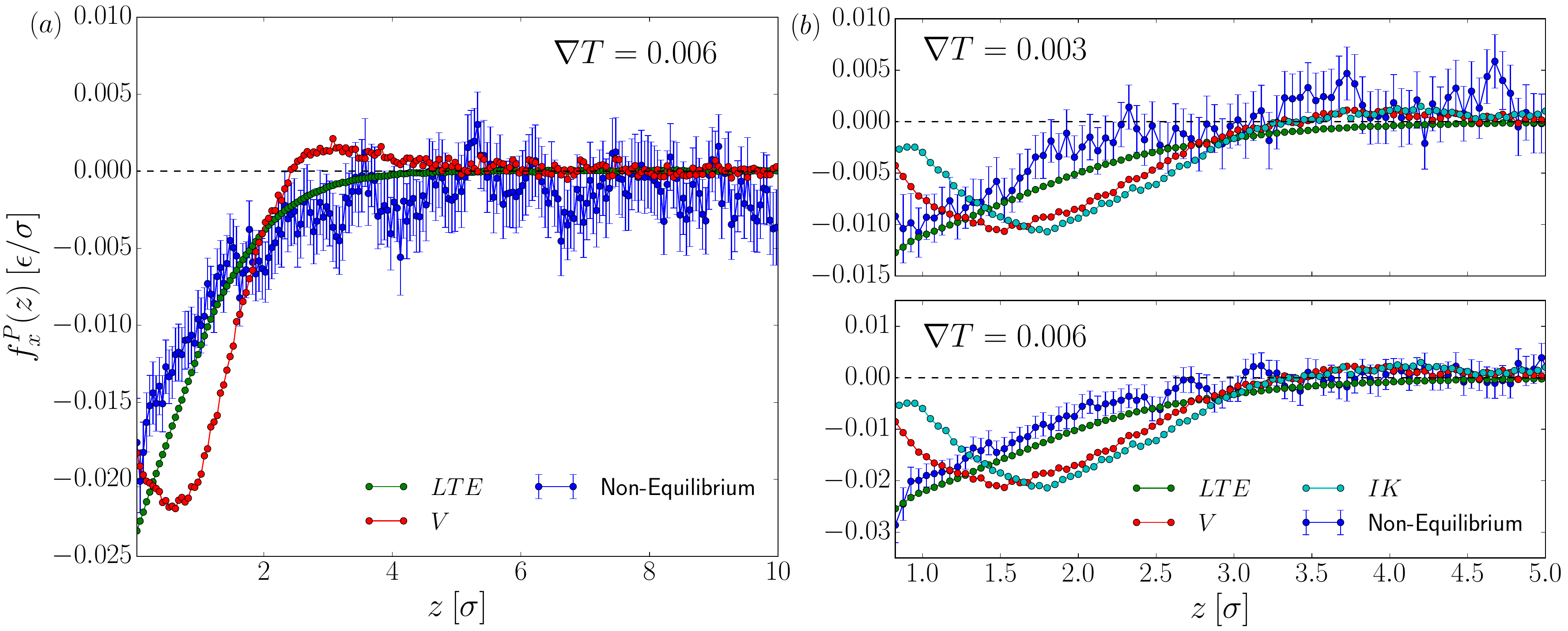}
\caption{Comparison of non-equilibrium force measurement (blue) with `stress gradient' approaches (cyan, red) and LTE approach (green) for (a) flat, reflective wall and (b) WCA wall. Below (a) $z = 0.025$ and (b) $z = 0.825$, the fluid density is less than 10\% of the bulk density giving poor statistics.\label{fig:reflect_wca_forces}}
\end{center}
\end{figure*}

Fig~\ref{fig:reflect_wca_forces}(a, b) compare the force per particle predicted by the stress-gradient and LTE methods with those computed directly via the non-equilibrium technique. For the structured wall shown in Fig~\ref{fig:reflect_wca_forces}(b) (see Fig~S3 for Lennard-Jones surface), the non-equilibrium calculation was carried out for temperature gradients of different magnitudes in order to validate the signal. As expected, the thermo-osmotic force is a monotonically increasing function of the gradient. To improve statistics, the non-equilibrium forces from the left and right regions were averaged (see Fig~S2).

Surprisingly, in all cases, both the $V$ (red) and $IK$ (cyan) stress gradients fail to predict the thermo-osmotic force (blue). Moreover, results for the reflective wall in Fig~\ref{fig:reflect_wca_forces}(a) show that the discrepancy is not due to wall contributions to the stress gradient.

In our previous work, there was significant numerical evidence indicating that stress gradients fail to predict microscopic Marangoni flows due to concentration gradients. Irving and Kirkwood suggested that an interface can cause the stress to depend on gradients of the pairwise atomic density, which are higher order terms neglected in the standard Irving-Kirkwood expression (Eq\eqref{eq:irving_kirkwood}) in the absence of gradients. The present work provides further evidence that the problem hinted at by Irving and Kirkwood becomes important in a temperature gradient near an interface, as the potential component of the stress tensor then depends not only on the distance of two points between which a force acts, but also on the absolute coordinates of these points.

Perhaps more significantly, the LTE approach (green) gets quite close, but still differs from the non-equilibrium result (blue). It is possible that this difference is due to deviation of the non-equilibrium result from the local thermal equilibrium approximation. Encouragingly, all methods agree in predicting zero net force in the bulk, consistent with the theory (Eq~\eqref{eq:gradP_enthalpy}).
%\begin{figure}[tb]
%\includegraphics[width=1.0\columnwidth]{reflect_full_force_comparison_poster.pdf}
%\caption{Comparison of non-equilibrium force measurement (blue) with `stress gradient' approach (red) and LTE approach (green) for a reflective wall.\label{fig:reflective_forces}}
%\end{figure}

As mentioned earlier, a flat solid wall should exert zero net transverse force $f_{x}^{P}$ on a fluid atom even in the presence of a thermal gradient. Yet in the case of an atomically structured wall, the stress gradient predicts a non-zero force contribution from the surface. To examine more deeply the role of the wall stress, we modified our non-equilibrium force calculation. Instead of including all intermolecular forces, we summed over only wall-fluid interactions $\phi_{wf}$ (see Supplementary Material: Wall Stress). In the case of a structured surface that has an attractive component, the wall does indeed exert an osmotic force (Fig S5(b)). The wall force can be attributed to the density gradient in $x$ shifting the average fluid atom positions so that they are asymmetric with respect to the position of solid atoms below (Fig S5(a)). Surprisingly, the stress gradient method (Fig S5(c)) predicts wall forces that are opposite in sign to the actual values. It is likely in the case of a purely repulsive surface, fluid will on average be sufficiently far away such that the latter force will become exceedingly small.

In this paper,  we have reported direct calculations of the thermo-osmotic force using a non-equilibrium simulation technique. We find that near a solid-fluid interface, $f_{x} \ne -\partial P_{xx}/\partial x$ suggesting that neither the Irving-Kirkwood nor virial expression accurately predict surface forces due to temperature gradients. Third, we find that an expression for the thermo-osmotic force based on the local enthalpy gets close to the true result. Finally, we have determined the contribution from wall stresses. For a structured surface, the wall structure does play a role in thermo-osmosis due to the asymmetric positioning of fluid atoms with respect to the lattice positions of solid atoms. This contribution disappears in the case of an unstructured reflecting wall.

This work was supported by the European Union grant 674979 [NANOTRANS]. We gratefully acknowledge numerous discussions with Lyd\'{e}ric Bocquet, Mike Cates, Patrick Warren, Ignacio Pagonabarraga and Benjamin Rotenberg. Additionally, we are grateful to Peter Wirnsberger for helpful suggestions with the method. RG gratefully acknowledges a PhD Grant from the Sackler Fund.

\bibliographystyle{apsrev4-1}
\bibliography{/Users/ramanganti/Aug25_Presentation/PRL_Submission/Primary_Manuscript/thermo_bib}

\begin{thebibliography}{10}%
\makeatletter
\providecommand \@ifxundefined [1]{%
 \@ifx{#1\undefined}
}%
\providecommand \@ifnum [1]{%
 \ifnum #1\expandafter \@firstoftwo
 \else \expandafter \@secondoftwo
 \fi
}%
\providecommand \@ifx [1]{%
 \ifx #1\expandafter \@firstoftwo
 \else \expandafter \@secondoftwo
 \fi
}%
\providecommand \natexlab [1]{#1}%
\providecommand \enquote  [1]{``#1''}%
\providecommand \bibnamefont  [1]{#1}%
\providecommand \bibfnamefont [1]{#1}%
\providecommand \citenamefont [1]{#1}%
\providecommand \href@noop [0]{\@secondoftwo}%
\providecommand \href [0]{\begingroup \@sanitize@url \@href}%
\providecommand \@href[1]{\@@startlink{#1}\@@href}%
\providecommand \@@href[1]{\endgroup#1\@@endlink}%
\providecommand \@sanitize@url [0]{\catcode `\\12\catcode `\$12\catcode
  `\&12\catcode `\#12\catcode `\^12\catcode `\_12\catcode `\%12\relax}%
\providecommand \@@startlink[1]{}%
\providecommand \@@endlink[0]{}%
\providecommand \url  [0]{\begingroup\@sanitize@url \@url }%
\providecommand \@url [1]{\endgroup\@href {#1}{\urlprefix }}%
\providecommand \urlprefix  [0]{URL }%
\providecommand \Eprint [0]{\href }%
\providecommand \doibase [0]{http://dx.doi.org/}%
\providecommand \selectlanguage [0]{\@gobble}%
\providecommand \bibinfo  [0]{\@secondoftwo}%
\providecommand \bibfield  [0]{\@secondoftwo}%
\providecommand \translation [1]{[#1]}%
\providecommand \BibitemOpen [0]{}%
\providecommand \bibitemStop [0]{}%
\providecommand \bibitemNoStop [0]{.\EOS\space}%
\providecommand \EOS [0]{\spacefactor3000\relax}%
\providecommand \BibitemShut  [1]{\csname bibitem#1\endcsname}%
\let\auto@bib@innerbib\@empty
%</preamble>
\bibitem [{\citenamefont {Lippmann}(1907)}]{lippmann1907}%
  \BibitemOpen
  \bibfield  {author} {\bibinfo {author} {\bibnamefont {Lippmann}},\
  }\href@noop {} {\bibfield  {journal} {\bibinfo  {journal} {C.R. Acad. Sci.}\
  }\textbf {\bibinfo {volume} {145}},\ \bibinfo {pages} {105} (\bibinfo {year}
  {1907})}\BibitemShut {NoStop}%
\bibitem [{\citenamefont {Aubert}(1912)}]{aubert1912}%
  \BibitemOpen
  \bibfield  {author} {\bibinfo {author} {\bibnamefont {Aubert}},\ }\href@noop
  {} {\bibfield  {journal} {\bibinfo  {journal} {Ann. Chim. Phys.}\ }\textbf
  {\bibinfo {volume} {26}},\ \bibinfo {pages} {551} (\bibinfo {year}
  {1912})}\BibitemShut {NoStop}%
\bibitem [{\citenamefont {Chretien}\ \emph {et~al.}(2017)\citenamefont
  {Chretien}, \citenamefont {Benit}, \citenamefont {Ha}, \citenamefont
  {Keipert}, \citenamefont {El-Khoury}, \citenamefont {Chang}, \citenamefont
  {Jastroch}, \citenamefont {Jacobs}, \citenamefont {Rustin},\ and\
  \citenamefont {Rak}}]{chretien2017mitochondria}%
  \BibitemOpen
  \bibfield  {author} {\bibinfo {author} {\bibfnamefont {D.}~\bibnamefont
  {Chretien}}, \bibinfo {author} {\bibfnamefont {P.}~\bibnamefont {Benit}},
  \bibinfo {author} {\bibfnamefont {H.}~\bibnamefont {Ha}}, \bibinfo {author}
  {\bibfnamefont {S.}~\bibnamefont {Keipert}}, \bibinfo {author} {\bibfnamefont
  {R.}~\bibnamefont {El-Khoury}}, \bibinfo {author} {\bibfnamefont
  {Y.}~\bibnamefont {Chang}}, \bibinfo {author} {\bibfnamefont
  {M.}~\bibnamefont {Jastroch}}, \bibinfo {author} {\bibfnamefont
  {H.}~\bibnamefont {Jacobs}}, \bibinfo {author} {\bibfnamefont
  {P.}~\bibnamefont {Rustin}}, \ and\ \bibinfo {author} {\bibfnamefont
  {M.}~\bibnamefont {Rak}},\ }\href@noop {} {\bibfield  {journal} {\bibinfo
  {journal} {bioRxiv}\ ,\ \bibinfo {pages} {133223}} (\bibinfo {year}
  {2017})}\BibitemShut {NoStop}%
\bibitem [{\citenamefont {Derjaguin}\ \emph {et~al.}(1987)\citenamefont
  {Derjaguin}, \citenamefont {Churaev},\ and\ \citenamefont
  {Muller}}]{derjaguin_surface}%
  \BibitemOpen
  \bibfield  {author} {\bibinfo {author} {\bibfnamefont {B.}~\bibnamefont
  {Derjaguin}}, \bibinfo {author} {\bibfnamefont {N.}~\bibnamefont {Churaev}},
  \ and\ \bibinfo {author} {\bibfnamefont {V.}~\bibnamefont {Muller}},\
  }\href@noop {} {\emph {\bibinfo {title} {Surface Forces}}}\ (\bibinfo
  {publisher} {Plenum, New York},\ \bibinfo {year} {1987})\BibitemShut
  {NoStop}%
\bibitem [{\citenamefont {Ganti}\ \emph {et~al.}(2017)\citenamefont {Ganti},
  \citenamefont {Liu},\ and\ \citenamefont {Frenkel}}]{ganti2017molecular}%
  \BibitemOpen
  \bibfield  {author} {\bibinfo {author} {\bibfnamefont {R.}~\bibnamefont
  {Ganti}}, \bibinfo {author} {\bibfnamefont {Y.}~\bibnamefont {Liu}}, \ and\
  \bibinfo {author} {\bibfnamefont {D.}~\bibnamefont {Frenkel}},\ }\href
  {\doibase 10.1103/PhysRevLett.119.038002} {\bibfield  {journal} {\bibinfo
  {journal} {Phys. Rev. Lett.}\ }\textbf {\bibinfo {volume} {119}},\ \bibinfo
  {pages} {038002} (\bibinfo {year} {2017})}\BibitemShut {NoStop}%
\bibitem [{\citenamefont {Schofield}\ and\ \citenamefont
  {Henderson}(1982)}]{schofield1982statistical}%
  \BibitemOpen
  \bibfield  {author} {\bibinfo {author} {\bibfnamefont {P.}~\bibnamefont
  {Schofield}}\ and\ \bibinfo {author} {\bibfnamefont {J.}~\bibnamefont
  {Henderson}},\ }\href@noop {} {\bibfield  {journal} {\bibinfo  {journal}
  {Proc. R. Soc. A}\ }\textbf {\bibinfo {volume} {379}},\ \bibinfo {pages}
  {231} (\bibinfo {year} {1982})}\BibitemShut {NoStop}%
\bibitem [{\citenamefont {Hansen}\ and\ \citenamefont
  {McDonald}(1990)}]{hansen1990theory}%
  \BibitemOpen
  \bibfield  {author} {\bibinfo {author} {\bibfnamefont {J.}~\bibnamefont
  {Hansen}}\ and\ \bibinfo {author} {\bibfnamefont {I.}~\bibnamefont
  {McDonald}},\ }\href@noop {} {\emph {\bibinfo {title} {Theory of Simple
  Liquids}}}\ (\bibinfo  {publisher} {Elsevier},\ \bibinfo {year}
  {1990})\BibitemShut {NoStop}%
\bibitem [{\citenamefont {Irving}\ and\ \citenamefont
  {Kirkwood}(1950)}]{irving1950}%
  \BibitemOpen
  \bibfield  {author} {\bibinfo {author} {\bibfnamefont {J.}~\bibnamefont
  {Irving}}\ and\ \bibinfo {author} {\bibfnamefont {J.}~\bibnamefont
  {Kirkwood}},\ }\href@noop {} {\bibfield  {journal} {\bibinfo  {journal} {J.
  Chem. Phys}\ }\textbf {\bibinfo {volume} {18}},\ \bibinfo {pages} {817}
  (\bibinfo {year} {1950})}\BibitemShut {NoStop}%
\bibitem [{\citenamefont {Weeks}\ \emph {et~al.}(1971)\citenamefont {Weeks},
  \citenamefont {Chandler},\ and\ \citenamefont {Andersen}}]{weeks1971role}%
  \BibitemOpen
  \bibfield  {author} {\bibinfo {author} {\bibfnamefont {J.}~\bibnamefont
  {Weeks}}, \bibinfo {author} {\bibfnamefont {D.}~\bibnamefont {Chandler}}, \
  and\ \bibinfo {author} {\bibfnamefont {H.}~\bibnamefont {Andersen}},\
  }\href@noop {} {\bibfield  {journal} {\bibinfo  {journal} {J. Chem. Phys}\
  }\textbf {\bibinfo {volume} {54}},\ \bibinfo {pages} {5237} (\bibinfo {year}
  {1971})}\BibitemShut {NoStop}%
\bibitem [{\citenamefont {Plimpton}(1995)}]{plimpton1995fast}%
  \BibitemOpen
  \bibfield  {author} {\bibinfo {author} {\bibfnamefont {S.}~\bibnamefont
  {Plimpton}},\ }\href@noop {} {\bibfield  {journal} {\bibinfo  {journal} {J.
  Comput. Phys.}\ }\textbf {\bibinfo {volume} {117}},\ \bibinfo {pages} {1}
  (\bibinfo {year} {1995})}\BibitemShut {NoStop}%
\end{thebibliography}%
     
\end{document}